\documentclass[%
 reprint,
 amsmath,amssymb,
 aps,
]{revtex4-2}

\usepackage{graphicx}
\usepackage{dcolumn}
\usepackage{bm}
\usepackage{xcolor} 
\usepackage{multirow}
\usepackage{comment}

\usepackage[colorlinks]{hyperref}
\hypersetup{%
	plainpages=true,
	breaklinks=true,
	hypertexnames=false,
	pageanchor=true,
	colorlinks=true,
	linkcolor={blue},
	citecolor={red},
	urlcolor={blue},
	anchorcolor={black}
}
\usepackage{siunitx} 
\usepackage[normalem]{ulem}
\usepackage{hyperref} 
\usepackage{cleveref} 
\usepackage{lineno}

\newcommand{\figpanel}[2]{Fig.~\hyperref[#1]{\ref*{#1}(#2)}}
\newcommand{\figpanels}[3]{Fig.~\hyperref[#1]{\ref*{#1}(#2-#3)}}
\newcommand{\figpanelNoPrefix}[2]{\hyperref[#1]{\ref*{#1}(#2)}}
\newcommand{\fullfigpanel}[2]{Figure~\hyperref[#1]{\ref*{#1}(#2)}}
\newcommand{\fullfigpanels}[3]{Fig.~\hyperref[#1]{\ref*{#1}(#2, #3)}}
\newcommand{\RNum}[1]{\uppercase\expandafter{\romannumeral #1\relax}}

\begin{document}

\preprint{APS/123-QED}

\title{Entanglement of photonic modes from a continuously driven two-level system} 

\author{Jiaying Yang$^{1,2}$}
\email{jiyang@chalmers.se}
\author{Ingrid Strandberg$^{1}$}
\author{Alejandro Vivas-Viaña$^{3,4}$}
\author{Akshay Gaikwad$^{1}$}
\author{Claudia Castillo-Moreno$^{1}$}
\author{Anton Frisk Kockum$^{1}$}
\author{Muhammad Asad Ullah$^{2}$}
\author{Carlos Sánchez Muñoz$^{3,4}$}
\author{Axel Martin Eriksson$^{1}$}
\author{Simone Gasparinetti$^{1}$}
\email{simoneg@chalmers.se}
\homepage{https://202q-lab.se}

\address{$^1$Department of Microtechnology and Nanoscience, Chalmers University of Technology, SE-412 96, G\"{o}teborg, Sweden
\\$^2$  Ericsson Research, Ericsson AB, SE-164 83, Stockholm, Sweden
\\$^3$ Departamento de Física Teórica de la Materia Condensada and Condensed Matter Physics Center (IFIMAC),
Universidad Autónoma de Madrid, 28049 Madrid, Spain 
\\$^4$  Institute of Fundamental Physics IFF-CSIC, Calle Serrano 113b, 28006 Madrid, Spain
}

\date{\today}             

\begin{abstract}
The ability to generate entangled states of light is a key primitive for quantum communication and distributed quantum computation. Continuously driven sources, including those based on spontaneous parametric downconversion, are usually probabilistic, whereas deterministic sources require accurate timing of the control fields. Here, we experimentally generate entangled photonic modes by continuously exciting a quantum emitter – a superconducting qubit – with a coherent drive, taking advantage of mode matching in the time and frequency domain. Using joint quantum state tomography and logarithmic negativity, we show that entanglement is generated between modes extracted from the two sidebands of the resonance fluorescence spectrum. Because the entangled photonic modes are perfectly orthogonal, they can be transferred into distinct quantum memories. Our approach can be utilized to distribute entanglement at a high rate in various physical platforms, with applications in waveguide quantum electrodynamics, distributed quantum computing, and quantum networks.
\end{abstract}

\maketitle

\section{Introduction}
Entanglement is a fundamental property of quantum physics, describing nonlocal correlations that are paramount for secure quantum communications~\cite{Bostrom2002,Lloyd2003,Pennacchietti2024}, remote quantum sensing~\cite{Croke2012, Lukin2019, Smith2023}, quantum algorithms~\cite{Jozsa2003, Nakhl2024}, and large-scale distributed quantum computing~\cite{Kim2014, Rempe2021, Akhtar2023}. Traditionally, in quantum optics, entangled photons have been produced from spontaneous parametric downconversion in combination with beamsplitters and photodetectors for heralding~\cite{Kwiat1995Dec,Ramelow2013Mar}. The probabilistic nature of this process can be inconvenient, and more recent experiments have been able to generate entangled photonic states on demand, i.e., deterministically, with emitted photonic quantum bits (qubits) defined either in the polarization~\cite{Muller2014,Istrati2020,Thomas2022} or a time-bin~\cite{Besse2020,Knaut2024,Ferreira2024} degree of freedom. 

The time-frequency degree of freedom can be a valuable tool for high-dimensional quantum information processing, and optical time- or frequency-bin entangled two-photon states are typically produced in parametric downconversion or spontaneous four-wave mixing processes~\cite{Huy2010, Zhang2021Aug, Sabattoli2022Dec,Clementi2023Jan}. 
In these parametric processes, the temporal mode shape of the output is selected by the spectral shape of the pump~\cite{Ansari2018May}. In the microwave frequency range, frequency-entangled states have been produced via Josephson junction circuit elements~\cite{Eichler2011Sep,Delsing2020,gasparinetti2017correlations, Peugeot2021Jul, Esposito2022Apr,Perelshtein2022Aug,Jolin2023}, but besides considering time-bin qubit encoding~\cite{Kurpiers2019Oct, Ilves2020Apr}, the temporal degree of freedom has been overlooked.

Here, we demonstrate a simple scheme to generate entangled time-frequency bosonic states from the steady-state emission of a single quantum emitter,  a transmon-type superconducting circuit coupled to a waveguide.
When the emitter is driven close to resonance, it exhibits resonance fluorescence~\cite{kimble1976theory, astafiev2010resonance}, a cornerstone of quantum optics and a source of antibunched photons.
Studies of frequency-filtered modes of the emission spectrum have unveiled a rich landscape of multi-photon processes, evidencing the generation of non-classical correlations~\cite{UlhaqCascadedSinglephoton2012,PeirisTwocolorPhoton2015,PeirisFransonInterference2017,LopezCarrenoPhotonCorrelations2017,ZubizarretaCasalenguaConventionalUnconventional2020, LopezCarrenoEntanglementResonance2024}.
In the time domain, it has been predicted that under certain conditions, selected temporal modes from the resonance-fluorescence emission exhibit a negative Wigner function, a hallmark of nonclassicality~\cite{StrandbergNumericalStudy2019, QuijandriaSteadyStateGeneration2018}, and this prediction has been experimentally verified~\cite{LuPropagatingWignerNegative2021}. 

In the present study, we combine the time and frequency dimensions by selecting two temporally overlapping, but spectrally orthogonal photonic modes.
We provide evidence of entanglement through joint quantum state tomography of the two selected modes, from which we determine the logarithmic negativity as a measure of entanglement~\cite{PhysRevLett.95.090503}. For optimally chosen parameters, we show entanglement between the two modes with a logarithmic negativity of 0.062.
The generated entangled photonic modes could be physically extracted and transferred to quantum memories to perform quantum information processing tasks, entanglement distribution, or entanglement distillation~\cite{Salari2024, ZHANG2018112,Kurochkin2014}.
The demonstrated method is agnostic to the physical platform and can be extended to consider emission from quantum systems with different level diagrams and pumping schemes. 
Our results thus open a new avenue to extract entanglement from continuously driven quantum systems.

\section{Results}
\subsection{Implementation with a superconducting circuit}
\begin{figure}\centering
\includegraphics[width=0.95\linewidth]{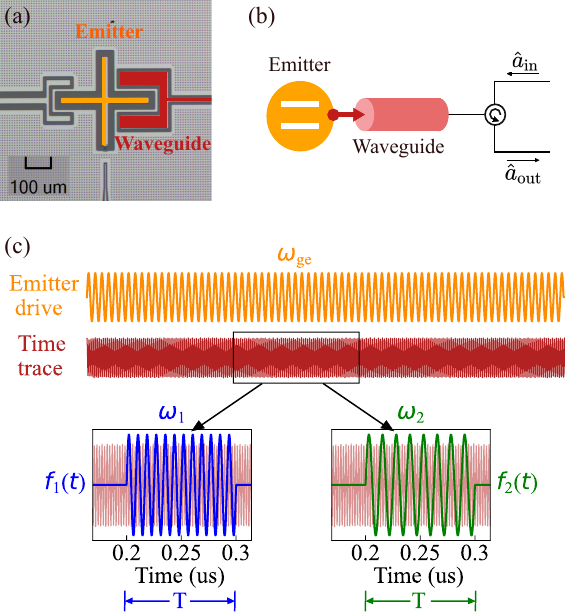}
\caption{Experimental implementation for entangled-photon generation based on a superconducting circuit. (a) False-color optical micrograph of the device. A transmon qubit (orange) is capacitively coupled to a waveguide (red). The coupling element visible to the left of the transmon is not used in this work.
(b) Schematic representation of the device and the measurement setup. $\hat a_{\rm in}$ and $\hat a_{\rm out}$ represent the input and output modes of the emitter's field, respectively. (c) Temporal mode matching. The emitter is continuously driven at its frequency $\omega_{\rm ge}$ (orange), and its emitted radiation is recorded as a time trace (red).
The two insets represent the temporal filters $f_1(t)$ (blue) and $f_2(t)$ (green) applied on the time trace to match the two propagating modes. 
}
\label{sample}
\end{figure}

We utilize an X-mon-type~\cite{barends2013coherent} superconducting circuit capacitively coupled to a waveguide [\figpanel{sample}{a}] as the quantum emitter. 
The transition frequency between the ground state and the first excited state of the emitter is $\omega_{\rm ge}/2\pi = \SI{4.94}{GHz}$.
The relaxation rate of the emitter into the waveguide is $\Gamma/2\pi=\SI{8}{MHz}$, corresponding to a relaxation time $T_1=1/\Gamma\sim\SI{20}{ns}$. 
The waveguide connects to the input and output lines in a reflection configuration [\figpanel{sample}{b})]. This setup guides the input field to the qubit via the input line and the waveguide. Subsequently, the reflected output field travels in the reverse direction through the waveguide, and we then measure it from the output line using a linear amplification chain (see Supplementary Material Sec.~\RNum{1} for more details on the setup). 
In the experiment, we continuously drive the emitter and measure the propagating output field in the waveguide [\figpanel{sample}{c}]. The drive has a strength characterized by the Rabi frequency $\Omega$ (see Supplementary Material Sec.~\RNum{4} for drive-strength calibration) and is on resonance with the fundamental transition of the emitter.

\subsection{Defining the temporal modes}

We apply two temporal filters $f_k(t)$ ($k=1,2$) to the output field, extracting single propagating modes $\hat{a}_k$ out of the time-dependent output field $\hat{a}^{\rm out} (t)$ as~\cite{LoudonQuantumTheory2000}
\begin{equation}
    \label{eq_TM}
        \hat{a}_k =\int_{-\infty}^{\infty} dt f_k(t) \hat a_{\text{out}}(t),
\end{equation}
The extracted modes fulfill bosonic commutation relations $[\hat a_k,\hat  a_k^\dagger]=1$, ensured by the normalization condition for the filter function $\int_{-\infty}^\infty dt |f_k(t)|^2=1$. Here,  $\hat a_{\text{out}}(t)$ is given by the input-output relation~\cite{LoudonQuantumTheory2000,GardinerQuantumNoise2004} $ \hat a_{\text{out}}(t)=\sqrt{\Gamma}\hat \sigma(t)-\hat a_{\text{in}}(t)$, where $\hat a_{\text{in}}(t)$ is the input field. 

Generally, it is possible to consider correlations between any pair of the propagating modes, including non-orthogonal ones. However, the orthogonality of the two temporal modes ensures that they are independent, and as such, can be physically extracted and mapped into separate quantum memories without added noise. For this reason, in the following, we will study correlations also for overlapping modes, but restrict to orthogonal modes when discussing entanglement (see Section~\ref{sec-2D} for more details).
The condition for the two temporal filters to be orthogonal is
\begin{equation}\label{eq:orthogonality}
\int f_1^*(t)f_2(t) dt=0 \, .
\end{equation}
where $f_1(t)$ and $f_2(t)$ are the two temporal filters with carrier frequencies $\omega_1$ and $\omega_2$ that fulfill
\begin{equation}\label{eq:temporal_filter}
f_k(t)=v_k(t)\cdot e^{i(\omega_k t+\phi)}, \quad k=1,2. 
\end{equation}
Here, $v_k(t)$ denotes the wavepacket profile of the $k$th filter. We take both profiles to be identical boxcar functions of duration $T$, such that $v_1(t) = v_2(t) = v(t)$.
For our case with two filters of the form Eq.~\eqref{eq:temporal_filter}, the orthogonality condition reduces to $(\omega_2 - \omega_1)/2\pi =  m/T$, where $m$ is an arbitrary integer. In this work, we use $T= \SI{100}{\ns}$, so the condition for orthogonality between the two modes is $(\omega_2 - \omega_1)/2\pi =  m\cdot \SI{10}{\MHz}$. Additionally, modes that are separated by a large frequency detuning can also be considered orthogonal to a good extent.

To characterize the modes $\hat a_k$, we apply the corresponding filter functions to the measured time traces and collect enough statistics to calculate the relevant moments of the distribution. To characterize and remove the added noise from the amplification chain, we interleave measurements with the drive on and off, while maintaining the same settings for the temporal filters, and apply known techniques~\cite{eichler2011a,eichler2012characterizing} to deconvolve the probability distribution of the output field from the added noise. 
Additionally, the coherent background, originating from the reflected input, is removed during post-processing (see Methods~\ref{remove_background} for details). To justify this operation, we note that it can be physically implemented without significant degradation of the modes, for example, using a cancellation tone fed into the output via a directional coupler~\cite{lu2021quantum}.

\subsection{Optimization of parameters for a single mode}
 \label{sec-1D}

\begin{figure}\centering
\includegraphics[width=0.95\linewidth]{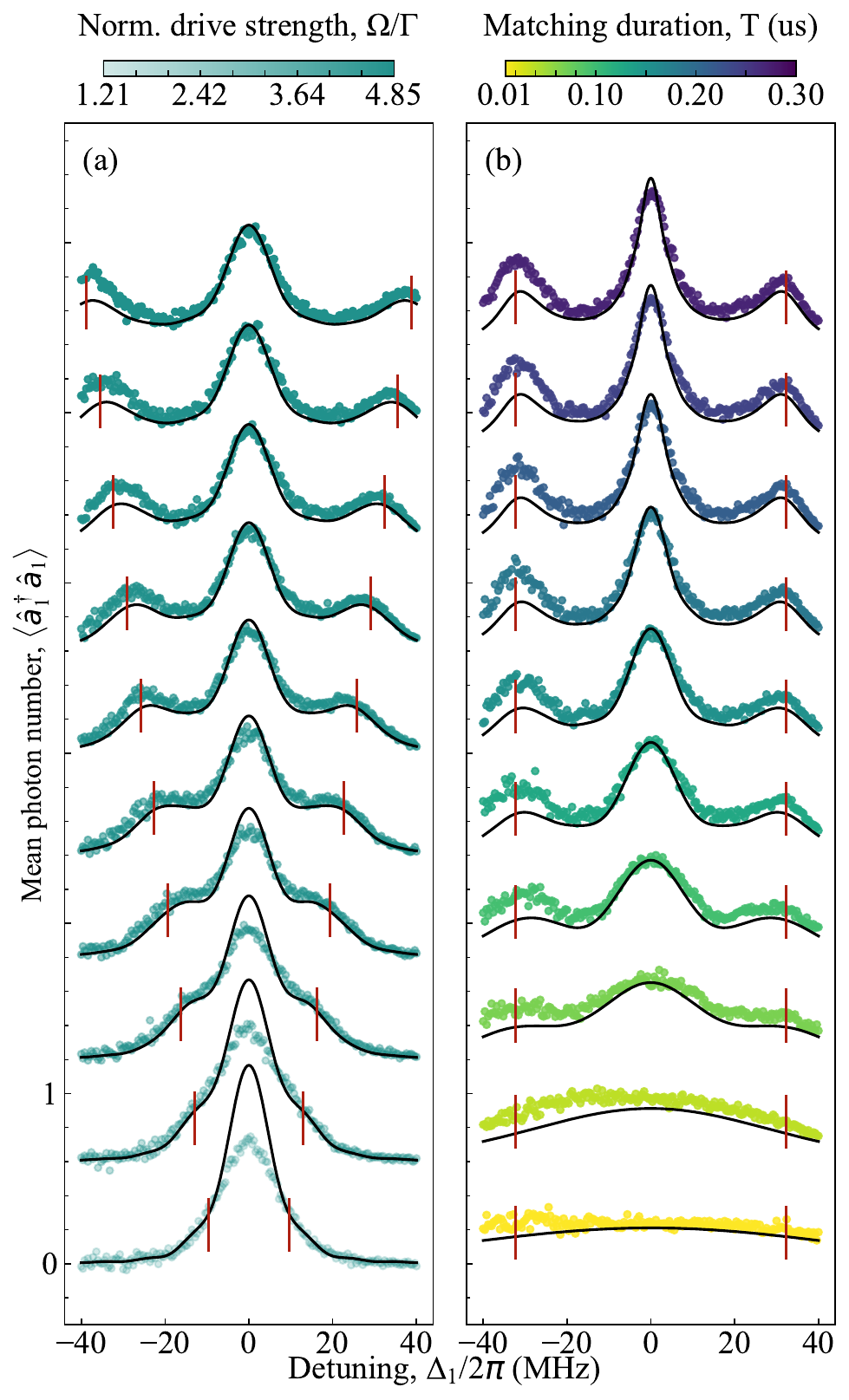}
\caption{Second-order moment $\langle \hat{a}_1^\dagger \hat{a}_1 \rangle$ of a single propagating mode under varying drive conditions. (a) \(\langle \hat{a}_1^{\dagger} \hat{a}_1 \rangle\) of the output mode as a function of Rabi frequency \(\Omega\), with the template-matching duration fixed at \(T = 100\,\text{ns}\). 
(b) \(\langle \hat{a}_1^{\dagger} \hat{a}_1 \rangle\) as a function of template-matching duration \(T\), with \(\Omega\) fixed at \(4.04\Gamma\). 
In both panels, each curve is vertically offset by 0.6 for clarity. The filled circles are the measured data, while the black solid lines are the simulation results. The red vertical lines mark the positions of $\pm \Omega$, which match the location of the side peaks of the corresponding curves.  
}
\label{fig-1D}
\end{figure}
We first characterize the power emitted into a single mode, $f_1(t)$.
We sweep the detuning $\Delta_1=\omega_1-\omega_{ge}$ between the modulation frequency of the temporal filter and the frequency of the emitter in the range of \qtyrange{-40}{40}{\MHz},
and measure the second-order moment $\langle \hat{a}_1^\dagger \hat{a}_1\rangle$, the mean photon number (Fig.~\ref{fig-1D}). 
With increasing drive power, side peaks appear at $\Delta_\pm \equiv \pm \Omega$~[\figpanel{fig-1D}{a}]. The frequency difference between the central peak and these side peaks corresponds to the drive Rabi frequency $\Omega$. 
This observed structure is the well-known Mollow triplet~\cite{MollowPowerSpectrum1969, astafiev2010resonance} of the resonance-fluorescence emission spectrum. In~\cite{LopezCarrenoEntanglementResonance2024}, Lopez et al.\ theoretically propose a method for generating entangled photons by off-resonantly driving a two-level system and measuring the emission from the side peaks of the Mollow triplet. In contrast to the continuous mode analysis  in~\cite{LopezCarrenoEntanglementResonance2024}, in the rest of the paper, we focus on exploring the entanglement of temporally selected modes at the side-peak frequencies while coherently driving the qubit.

The duration of the filter affects its spectral content~[\figpanel{fig-1D}{b}]; when the filter duration is short, the side peaks are not visible due to spectral broadening.
We model our measurements with master equation and input-output theory (see Methods~\ref{model}) and find a good agreement between theoretical predictions and experimental data.

In the following, we set the drive Rabi frequency to $\Omega = 4.04\Gamma$ and the duration of the temporal modes to $T = \SI{100}{ns}$, corresponding to a well-developed Mollow triplet and a filter function for which the side peaks are well resolved. This sideband-resolved regime allows us to obtain entanglement between the two orthogonal modes, as demonstrated in the next section.

\subsection{Two-mode entanglement}
 \label{sec-2D}
  
\begin{figure*}
\includegraphics[width=\linewidth]{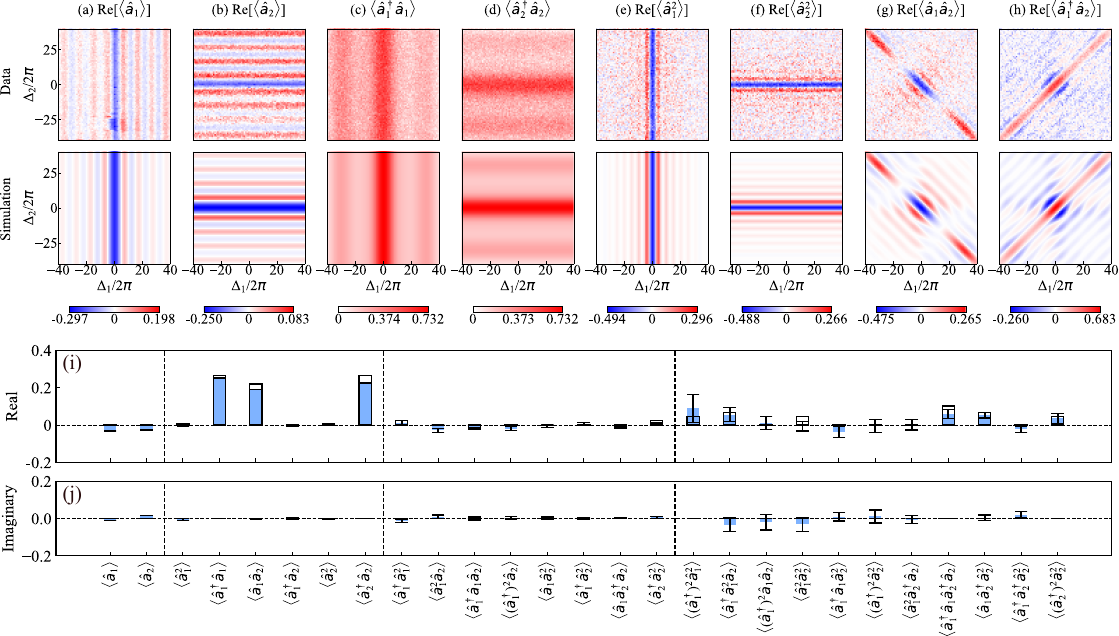}
\caption{Moments and correlations of temporally matched modes. (a-h) Moments up to second order for the temporal modes defined in Eq.~\eqref{eq:temporal_filter} as a function of frequency detunings $\Delta_k=\omega_k-\omega_{\rm ge}$, where $k=1,2$.  The upper row shows the data measured from the experiments. The lower row shows the 2D maps of the moments calculated from the simulation. The 2D map of the second-order moments $\langle \hat{a}_1^\dagger \hat{a}_1 \rangle$ and $\langle \hat{a}_2^\dagger \hat{a}_2 \rangle$ are normalized to have the same maximum value as the simulation, and the other moments are scaled accordingly. (i, j) The real and imaginary parts of moments up to fourth order at the frequencies $(\Delta_1,\Delta_2)=(\Delta_-, \Delta_+)$. The blue bar is the measured data, while the black wireframe is the simulation. The error bar for each measured moment is obtained by splitting the data into 20 segments, calculating the moments for each segment and obtaining the standard deviation over them.  }
\label{2D}
\end{figure*}

\begin{figure}
\centering
\includegraphics[width=0.9\linewidth]{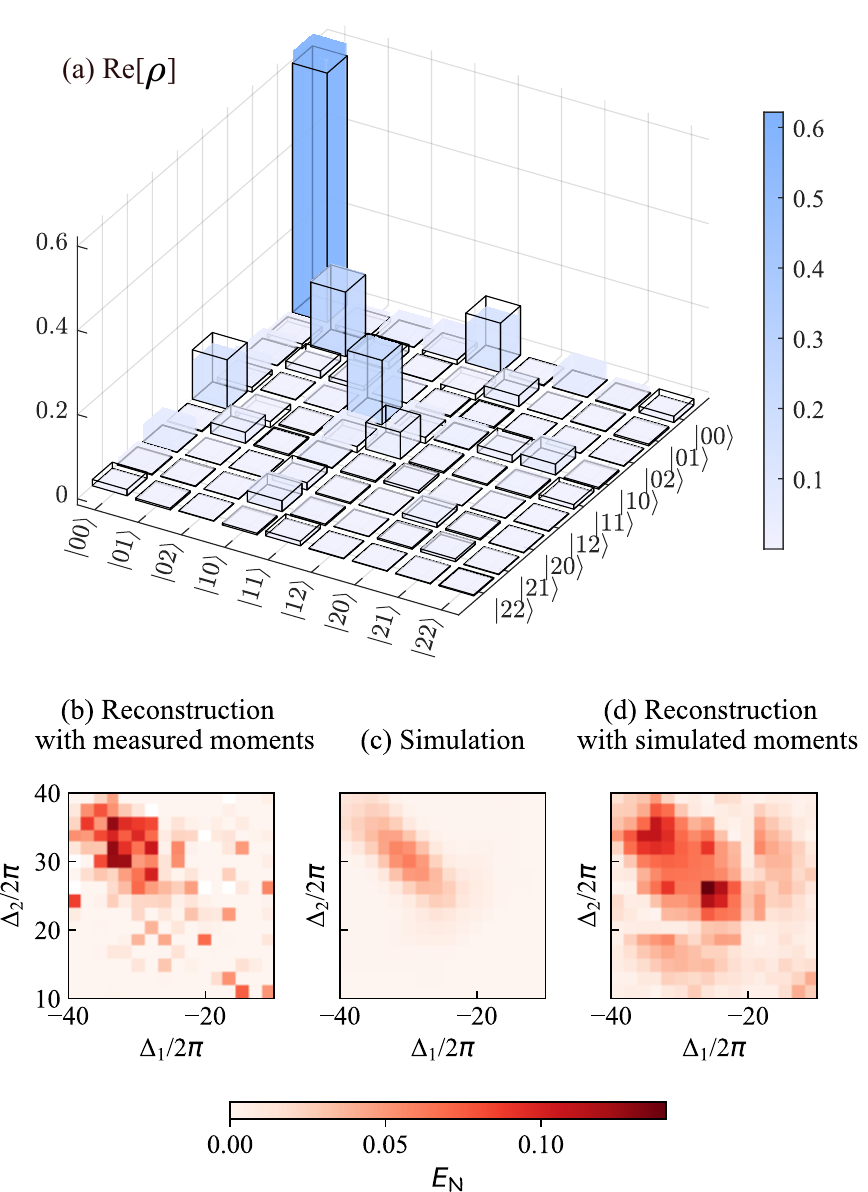}
\caption{Entangled temporal modes.  Density matrix from the joint quantum state tomography and the obtained logarithmic negativity. (a) Comparing the real part of the reconstructed and simulated density matrices in Fock space up to $N=5$, while only the selected lower-order components satisfying $N\le3$ are shown. The colored bars show the measurement result and the black wireframes are the simulated prediction. The imaginary components of both the simulated and reconstructed density matrices, which are not shown, are less than 0.015 across all elements of the matrices. 
(b) The logarithmic negativity $E_\mathcal{N}$ reconstructed from the 27 measured moments over the frequency ranges $\Delta_1 \in [-40, -10]\,\mathrm{MHz}$ and $\Delta_2 \in [10, 40]\,\mathrm{MHz}$. There are seven not-a-number points, shown in white, due to the optimizer failing to meet the standard deviation constraints of the moments at these frequency points during reconstruction. (c) The numerically simulated logarithmic negativity $E_\mathcal{N}$ over the same frequency range. (d) $E_\mathcal{N}$ reconstructed from the 27 simulated moments over the same frequency ranges. This distribution matches (c) if reconstructed using all combinations of simulated moments up to $N=5$ (see Methods~\ref{Methods-QST}).
}
\label{tomography}
\end{figure}

We match the measured time trace with two simultaneously applied temporal filters, $f_1(t)$ and $f_2(t)$, to verify entanglement at optimal frequencies between the two selected propagating modes. The temporal profiles of the filters $v(t)$ are perfectly overlapping, $v_1(t) = v_2(t)$, while their frequencies, $\omega_1$ and $\omega_2$, are independently varied in the range [$\omega_{\rm ge}-\SI{40}{MHz},\omega_{\rm ge}+\SI{40}{MHz}$]. 
At each frequency of both temporal filters, we obtain the first- and second-order moments of the two propagating modes [See \figpanels{2D}{a}{h}. The first row shows the simulated results (from the model in Methods~\ref{model}) and the second row shows the measured results (See Methods~\ref{Methods-denoise} to~\ref{Mehods-twomode} for data-analysis details)]. In the frequency regions around $(\Delta_1,\Delta_2)=(\Delta_-, \Delta_+)$ or $(\Delta_+, \Delta_-)$, near the anti-diagonal corners of the 2D maps, the  moments $\langle \hat a_1 \rangle$ and $\langle \hat a_2 \rangle$ are close to zero [\fullfigpanels{2D}{a}{b}], while the cross-second-order moment $\left \langle \hat a_1 \hat a_2 \right \rangle$ shows a peak [\figpanel{2D}{g}].
This indicates a two-mode squeezing type of entanglement. 
While in the region near $(\Delta_1,\Delta_2)=(\Delta_+, \Delta_+)$ or $(\Delta_-, \Delta_-)$, along the diagonal of the 2D moment maps, the cross-second-order moment $\langle \hat a_1 ^\dagger \hat a_2 \rangle$ shows a peak [\figpanel{2D}{h}], indicating a beam-splitter type of entanglement.
Note, however, that in this scenario, the two modes overlap both temporally and in frequency, and are generally not orthogonal.
In the following, we focus on the frequency point $(\Delta_-, \Delta_+)$, where the two modes belong to opposite side peaks of the Mollow triplet and satisfy the frequency orthogonality condition in Eq.~\eqref{eq:orthogonality}.
At the selected point, we reduce the measurement noise level by performing more repetitions ($n$ = $2\times 10^7$), allowing for the computation of moments up to the fourth order. Specifically, we compute moments $ \langle (\hat{a}_1^{\dagger})^{m_1} \hat{a}_1^{m_1} (\hat{a}_2^{\dagger})^{m_2} \hat{a}_2^{n_2} \rangle$ for $m_1, n_1, m_2, n_2 \in \left \{  0, 1, 2\right \}$ and 
$m_1+n_1+m_2+n_2 \le 4$, resulting in 27 moments excluding conjugation redundancy [\fullfigpanels{2D}{i}{j}]. 

At this frequency point, we also reconstruct the density matrix of the two propagating modes by joint quantum state tomography~[\figpanel{tomography}{a}]. The tomography utilizes least-squares optimization~\cite{PhysRevResearch.2.042002, PhysRevApplied.18.044041} to find the most-likely density matrix corresponding to the measured
moments [\figpanel{tomography}{a}].
In the reconstructed density matrix, multiple photon states are involved. Although the component $\langle 00 | \rho | 11 \rangle$ (where $|ij\rangle$ represents the number of photons in each mode) is the largest, there are additional off-diagonal elements (coherences) that are nonzero. The state overlap between the reconstructed and the simulated density matrices is 96.6\%, with the state overlap defined as~\cite{jozsa1994fidelity}
$F(\rho, \rho') = \left(\mathrm{Tr}\sqrt{\sqrt{\rho}\rho'\sqrt{\rho}}\right)^2$,
where $\rho$ and $\rho'$ are the simulated and measured density matrix, respectively.  

Further, to demonstrate that the two modes are entangled around the selected frequency point  $(\Delta_-, \Delta_+)$, we perform joint quantum state tomography with $n$ = $2\times 10^7$ repetitions across the frequency ranges $\Delta_1 \in [-40, -10]\,\mathrm{MHz}$ and $\Delta_2 \in [10, 40]\,\mathrm{MHz}$. We then calculate the logarithmic negativity $E_\mathcal{N}$ from the reconstructed density matrices as a metric to quantify entanglement. 
We notice that the computation of $E_\mathcal{N}$ is very sensitive to small values of coherences of the density matrix. As a result, its value presents large fluctuations when we use least-squares optimization to reconstruct the density matrix.
To reduce fluctuations in $E_\mathcal{N}$, we employ compressed-sensing optimization~\cite{shabani-prl-2011, rodi-prb-2014}, which is a more aggressive approach that aims to find the optimal sparse solution of the density matrices. Given that our target density matrix is sufficiently sparse and we utilize a heavily reduced data set [using only 27 moments up to the fourth order, listed in \fullfigpanels{2D}{i}{j}, for reconstruction], compressed-sensing optimization effectively minimizes the impact of noisy coherent components in the reconstructed density matrices. This method thereby enables a clearer distribution of $E_\mathcal{N}$ [see \figpanel{tomography}{b}, and see Methods~\ref{Methods-QST} for details of the two optimization methods].

We observe the maximum $E_\mathcal{N}$ from the reconstruction near the frequency point $(\Delta_-, \Delta_+)$, with $E_\mathcal{N}=0.128$, while that from the simulation reaches a maximum value of 0.062, at the same point
[\figpanel{tomography}{c}] (See Supplementary Material Sec.~\RNum{2} for simulations in more parameter regimes).
We conjecture that the discrepancy between the simulated and measured $E_\mathcal{N}$ is due to our use of 27 moments up to fourth order, instead of the full 325 moments for a Fock-space cutoff at $N=5$, which includes all cases fulfilling $m_1, n_1, m_2, n_2 \in \left \{  0, 1, 2, 3, 4\right \}$. The excluded moments are between the fifth and 16th order, which are omitted due to high noise levels. Indeed, if we reconstruct the state using the same moments as done for the experiment, using simulated values, we obtain consistently higher values for $E_\mathcal{N}$, very close to the experiment~[\figpanel{tomography}{d}].
By contrast, if we use all 325 combinations of simulated moments, we reproduce the  $E_\mathcal{N}$ distribution obtained directly from the simulation (see direct comparison in  Methods~\ref{Methods-QST}). Based on these considerations, we conclude that the maximum $E_\mathcal{N}$ observed in our experiment is close to the simulation value of 0.062.

\section{Discussion}
\label{Conclu}

Our work introduces and demonstrates an approach for generating entanglement in the time-frequency domain between propagating bosonic modes. 
Theoretical studies suggest that entangled photons are generated in the Mollow triplet sideband emission observed in the resonance fluorescence from a single two-level emitter~\cite{SanchezMunozViolationClassical2014,LopezCarrenoEntanglementResonance2024}.
We experimentally confirm generation of entanglement from steady-state resonance fluorescence in a continuously and coherently driven emitter by temporally matching two photonic modes from the continuous spectrum.
At the frequencies of the opposite side peaks of the Mollow triplet, the reconstructed density matrix of the two-mode state exhibits a maximum logarithmic negativity of 0.128, which provides evidence of entanglement, aligning with our theoretical model of 0.062 after considering the limitations of our reconstruction scheme. 

Entanglement in the time-frequency degree of freedom is a valuable asset for high-dimensional quantum information processing. In previous experiments, time-frequency modes have been utilized with entangled single-photon pairs~\cite{Ansari2018May,Ashby2020Dec,Chiriano2023Oct}. Our approach naturally incorporates multiphoton states, in which the number of photons is determined by the mode shape and output intensity of the quantum emitter. This experiment used a boxcar filter to define temporal modes, but simulation results presented in Supplementary Material~\RNum{2} suggest that Hermite--Gauss modes yield higher logarithmic negativity. As such, there is potential to explore and optimize the mode shape to obtain states with higher entanglement or other desired properties.

While in our experiment we selected and characterized the modes simply by applying digital filters, these modes could be physically extracted and transferred to quantum memories~\cite{Bao2021, Matanin2023} to perform quantum information processing tasks or entanglement distribution. In the optical regime, the so-called quantum pulse gate~\cite{Eckstein2011Jul} is capable of selecting an arbitrary time-frequency mode from a high-dimensional input~\cite{Ansari2018May,Harder2017Aug}. Selected modes can also be captured in cavities by tunable coupling~\cite{Xu2018Mar,Yin2013Mar}, a technique which is well suited for microwave photons in superconducting circuits, in which cavities with adjustable couplers can be fabricated on-chip.


Compared to previous works~\cite{wang2011deterministic, lang2013correlations, narla2016robust, gasparinetti2017correlations, kannan2020generating, Besse2020}, this system for entanglement generation has two main advantages: i) it uses a simple system with a single emitter; ii) the entangled modes are generated by driving the emitter at steady state, so there is no timing constraint in applying the temporal filters.
Additionally, the entanglement generation rate is only limited by the linewidth of the emitter, which can be made of the order of \SI{1}{GHz} in superconducting circuits~\cite{forn-diaz2017a}.

Although we focused on resonance fluorescence of a superconducting circuit, our method is broadly applicable since resonance fluorescence has been observed in a multitude of systems, including trapped ions~\cite{Zoller1993,Tamm2020}, cold atoms~\cite{Morigi2007}, color centers in diamond~\cite{Higbie2017,Chen2022, Kumar2021}, and semiconductor quantum dots~\cite{Muller2007,Makhonin2014}, even at telecom wavelength~\cite{Gerardot2016,Michler2021}. The method could also be applied using steady-state emission from other quantum systems as a resource, be it a comparatively simple arrangement such as a Kerr cavity~\cite{Strandberg2021}, or more complex systems, for example, emitters in topological waveguides~\cite{Mittal2018,Mehrabad2020Dec, Tudela2023} or those producing superradiance~\cite{Mlynek2014Nov}.


\section{Methods}

\subsection{Theoretical model}
\label{model}
\subsubsection{Entanglement source}
We describe the emitter as a two-level system, spanning a basis $\{|g\rangle,|e\rangle\}$, with transition frequency $\omega_{\rm ge}$ and lowering operator $\hat \sigma=|g\rangle \langle e|$. We focus on the case of resonant driving, where the emitter is excited by a microwave pulse of frequency $\omega_{\rm ge}$ and Rabi frequency $\Omega$. In a frame rotating at the drive frequency and under a rotating-wave approximation, the Hamiltonian of the driven qubit is given by 
\begin{equation}
    \hat H_{q}= -i\Omega(\hat \sigma^\dagger-\hat \sigma)/2\, .
\end{equation}

The interaction between the qubit and the coplanar waveguide introduces incoherent processes by which the qubit is de-excited through spontaneous emission at a rate $\Gamma$.
The dissipative dynamics of the reduced density matrix of the qubit is described by the master equation~\cite{BreuerTheoryOpen2007,CarmichaelOpenSystems1993,GarciaRipollQuantumInformation2022} 
\begin{equation}
    d \hat \rho/dt=-i[\hat H_q,\hat \rho]+\frac{\Gamma}{2}\mathcal{D}[\hat \sigma]\hat \rho\, ,
\end{equation} 
where we have defined the Lindblad superoperator as $\mathcal{D}[\hat A]\equiv 2 \hat A \hat  \rho \hat A^\dagger-\{\hat A^\dagger \hat A, \hat \rho\}$. 

A diagonalization of $\hat H_q$ yields two eigenstates $|\pm\rangle \equiv (|g\rangle \pm |e\rangle)/\sqrt{2}$ with corresponding eigenenergies $E_\pm = \pm \Omega$, which in the dressed-atom picture can be understood as hybrid light-matter states between the qubit and the drive. 
In the strong driving regime ($\Omega > \Gamma/4$), these eigenenergies can be resolved, and the emission spectrum of resonance fluorescence acquires the Mollow-triplet structure~\cite{MollowPowerSpectrum1969}, with two side peaks emerging around a central peak at the drive frequency (see Fig.~\ref{fig-1D}).
This structure can be understood as transitions between the dressed eigenstates, the central peak at $\omega = \omega_{\rm ge}$ corresponding to the doubly-degenerate transition $|\pm\rangle \rightarrow |\pm \rangle$, and the two side peaks at frequencies $\omega_{\pm}=\omega_{\rm ge}\pm \Omega$ corresponding to transitions $|+\rangle \rightarrow |-\rangle$ and $|-\rangle \rightarrow |+\rangle$, respectively.
\subsubsection{Joint quantum state of temporal modes}
\label{sec:cascaded_master_eq}

The density matrix describing the joint quantum states of both modes can be computed by using the input-output theory for quantum pulses~\cite{KiilerichInputOutputTheory2019,KiilerichQuantumInteractions2020}. This approach treats the system and the emission as a cascaded quantum system~\cite{GardinerQuantumNoise2004}, in which the desired temporal mode is described as a virtual cavity (with annihilation operator $\hat a_k$) coupled non-reciprocally to the system with a time-dependent coupling. Here, we extend this approach to capture simultaneously several temporal modes by introducing two virtual cavities, resulting in the following cascaded master equation in the rotating frame of the drive:
\begin{multline}
\label{eq:master_eq}
     \frac{d\hat \rho}{dt}=-i[\hat H_q,\hat \rho]+\frac{\Gamma}{2}\mathcal{D}[\hat \sigma]\hat \rho + \sum_{{k}=1}^2\frac{|g_k(t)|^2 }{2}\mathcal{D}[\hat a_{k}] \hat \rho
    \\
    -\sum_{{k}=1}^2 \sqrt{\Gamma } \left(g^*_k(t)[\hat a_{k}^\dagger,\hat \sigma \hat \rho]+g_k(t)[\hat \rho \hat \sigma^\dagger,\hat a_{k}]\right).
\end{multline}   

The temporal filters defined in Eq.~\ref{eq:temporal_filter} are encoded into  time-dependent coupling~\cite{NurdinPerfectSingle2016,GoughGeneratingNonclassical2015},
\begin{equation}
    \label{eq:TimeDependentCoupling}
    g_{{k}}(t)=-\frac{f_{k}(t)}{\sqrt{\int_0^t dt' |f_{k}(t')|^2}}.
\end{equation}
In our implementation, the profile $v(t)$ of the temporal filters (see Eq.~\ref{eq:temporal_filter}) is defined as a normalized boxcar function $ v(t)=\frac{1}{\sqrt{T}}[\Theta (t-t_0)-\Theta(t-t_0-T)]$, with $\Theta(t)$ the Heaviside step function and $t_0$ the start time of the temporal filters. 
This results in the time-dependent couplings (in the rotating frame of the drive) $g_k(t)=-e^{i(\Delta_k t+\phi)}/\sqrt{t-t_0}$, where $\Delta
_k \equiv \omega_k-\omega_{\text{ge}}$ is the $k$th sensor-laser detuning. 

We note that related previous works~\cite{LopezCarrenoFrequencyresolvedMonte2018,LopezCarrenoEntanglementResonance2024} used a version of the master equation in Eq.~\eqref{eq:master_eq} that includes a factor $1/\sqrt{2}$ in the last term. Such a factor would stem from a description in which, prior to the capture of each mode, the output is physically split (e.g. by a beam splitter). In that description, the state obtained for each mode depends strongly on the total number of modes included, since the physical splitting has the effect of introducing important vacuum contributions. This would not accurately describe the digital filtering of the modes performed in this experiment, where the particular choice of $f_1$ should not affect the results obtained when filtering $f_2$. This condition is not fulfilled by the cascaded multi-mode setups proposed in Ref.~\cite{KiilerichInputOutputTheory2019}. The master equation in Eq.~\eqref{eq:master_eq}, however, ensures independence of the filtered modes, and therefore serves as a good description of the type of mode-matching performed in this experiment.

Experimentally, $t_0 =200$ ns is a sufficient waiting time to reach steady state. Numerically, we evolve the system until time $t_0+T$ solving the time-dependent cascaded master equation Eq.~\eqref{eq:master_eq} using QuTiP~\cite{JohanssonQuTiPOpensource2012,JohanssonQuTiPPython2013, lambert2024qutip}. The simulated density matrix and moments of the two-mode state are obtained and used for comparison with experimental results.
Then, we quantify the degree of entanglement between the two filtered temporal modes by means of the logarithmic negativity~\cite{vidal2002computable,PlenioIntroductionEntanglement2007,HorodeckiQuantumEntanglement2009}, a commonly used entanglement witness in bipartite systems. Given a general bipartite state,
composed by systems $A$ and $B$, the logarithmic negativity is defined as
\begin{equation}
    E_\mathcal{N}\equiv\text{log}_2(||\rho^{\text{T}_A}||_1),
\end{equation}
where $\text{T}_A$ denotes the partial transpose operation oversystem $A$, and $||\cdot||_1$ is the trace norm. 

\subsection{Moment denoising}
\label{Methods-denoise}
Due to measurement noise arising from cable losses, amplification chain, mode-matching inefficiency, and other factors, the directly obtained total mode from temporal template matching, denoted as $\hat{S}_k$, does not solely represent the target modes $\hat{a}^{\rm out}_k$. Instead, $\hat{S}_k$ comprises both the target mode $\hat{a}_k^{\rm out}$ and an additional noise mode $\hat{h}^\dagger$, $\hat{S}_k = \hat{a}_k^{\rm out} + \hat{h}^\dagger$. To remove the noise mode $\hat{h}^\dagger$, we operate an interleaved measurement, to sweep between two cases with and without the qubit drive. In the first case, we measure the total mode including both the targeted mode and the noise mode. In the second case, the target mode is left in vacuum and the measurement can be served as a reference of the noise mode. The switching between the two cases is repeated $n$ times, with $n$ varying from $10^5$ to $2\times 10^7$ across different measurements, given that the added noise photon from the amplification chain is $n_{\rm added}$ = 11 in our setup. We use number of repetitions $n$ = $10^5$ in \figpanels{2D}{a}{h}, $n$ = $10^6$ in Fig.~\ref{fig-1D} and $n$ = $2\times 10^7$ in \fullfigpanels{2D}{i}{j} and Fig.~\ref{tomography}.  We then calculate the averaged moments from these repetitions. As we investigate higher-order moments, the number of required repetitions increases due to the escalating statistical errors associated with higher orders~\cite{da2010schemes}.

The first- and second-order moments of the two propagating modes~\cite{eichlor_thesis}, $\left<\hat{a}_k^{\rm out}\right>$ and $\left<(\hat{a}_{k}^{\rm out} )^\dagger\hat{a}_{k}^{\rm out}\right>$, are obtained by 
\begin{equation}
\begin{split}
\left<\hat{a}_k^{\rm out}\right> &= \left<\hat{S}_k-\hat{h}^\dagger\right>\, , \\
\left<(\hat{a}_{k}^{\rm out} )^\dagger\hat{a}_{k}^{\rm out}\right> &= \left<\hat{S}_{k}^\dagger\hat{S}_{k}-\hat{S}_{k}\hat{h}-\hat{S}_{k}^\dagger\hat{h}^\dagger+\hat{h}\hat{h}^\dagger\right>\, ,
\end{split}
\label{moments_cal_2D}
\end{equation}
where the angle brackets represent the averaging over the $n$ repetitions. These moments are with the coherent background from the reflected input drive; and in the next section, we discuss the subtraction of the coherent background. 

\subsection{Removal of coherent background}
\label{remove_background}

The relation between the input and the output modes is according to input-output theory~\cite{gardiner1985input}
\begin{equation}
\sqrt{\Gamma}\hat{\sigma}^-_k = \hat{a}^{\rm out}_k-\hat{a}^{\rm in}_k\, .
\end{equation}
The emission operator of the qubit, $\sqrt{\Gamma} \hat{\sigma}^-_k$, is the difference between the output field and the input field. 
In the main text, we choose to present experimental results and simulations for the emission operator of the qubit alone, that is, after the coherent input field has been removed. The subtraction of a coherent field of arbitrary amplitude and phase, corresponding to a displacement in phase space, is a physically justified operation. For example, in Ref.~\cite{lu2021quantum}, the subtraction is performed directly in the experimental setup, by adding a cancellation pulse through a weakly coupled directional coupler.
Here, we remove the coherent background from the output mode obtained through measurement by post-processing, instead.

We calculate the reflected input mode that is captured by the temporal filter through, 
\begin{equation}
\left<\hat{a}^{\rm in}_k\right> = \Omega/ \sqrt{\Gamma} \cdot \int_{0}^{T} e^{i(\omega_k t+\phi_B )} \cos(\omega_{\rm IF}t) dt\, ,
\label{input_mode}
\end{equation}
where $\Omega$ is the Rabi frequency of the drive signal, $\Gamma$ is the decay rate of qubit to the waveguide, $\omega_k$ is the frequency of the temporal filter, and $\omega_{\rm IF}$ is the down-converted frequency of the qubit. 

Aside from the phase rotation parameters of the input mode, $\phi_B$, we also introduce an amplification parameter \(A\) a phase rotation \(\phi_A\), to the measured output mode \(\hat{a}^{\text{out}}_k\), and an amplification parameter $B$ to the input mode \(\hat{a}^{\text{in}}_k\). The same set of parameters \(A\), \(B\), \(\phi_A\), and \(\phi_B\) are used for both modes.  By optimally selecting these parameters, we aim to equate the simulated first-order moment with the transformed combination of these fields, thus achieving parameter identification and system characterization:
\begin{equation}
\left<\hat{a}_k\right>_{\rm sim} = \left<Ae^{i\phi_A}\hat{a}^{\rm out}_k - B\hat{a}^{\rm in}_k\right>.
\end{equation}

Shifting our focus to the second-order moment, we relate simulation and measurement as follows:
\begin{equation}
\begin{split}
    \left<\hat{a}_k^\dagger \hat{a}_k\right>_{\rm sim} &= \left<(Ae^{i\phi_A} \hat{a}^{\rm out}_k - B \hat{a}^{\rm in}_k)^\dagger(Ae^{i\phi_A} \hat{a}^{\rm out}_k - B \hat{a}^{\rm in}_k)\right> \\
    &= A^2\left<(\hat{a}^{\rm out}_k)^\dagger \hat{a}^{\rm out}_k\right> + B^2\left<(\hat{a}^{\rm in}_k)^\dagger \hat{a}^{\rm in}_k\right>\\
    &\quad - AB\left(e^{-i\phi_A} \left<\hat{a}^{\rm in}_k (\hat{a}^{\rm out}_k)^\dagger\right> + e^{i\phi_A} \left<(\hat{a}^{\rm in}_k)^\dagger \hat{a}^{\rm out}_k\right>\right).
\end{split}
\label{eq-optimization}
\end{equation}

By using a Scipy~\cite{2020SciPy-NMeth} optimizer to align the measured moments [calculated with Eq.~(\ref{moments_cal_2D})] with the simulated second-order moment according to Eq.~(\ref{eq-optimization}), we find the optimal parameters $A'$, $\phi_A'$, $B'$, and $\phi_B'$. This procedure, including integration, optimization, and subtraction, is executed on both measured output modes. The optimal subtraction of the coherent background for each mode is 
\begin{equation}
\sqrt{\Gamma} \left< \hat{\sigma}^-_k \right>= A'e^{i\phi_A'}\left<\hat{a}^{\rm out}_k\right>-B'\left<\hat{a}^{\rm in}_k\right>\, .
\end{equation}
This method identifies the emission operator, $\sqrt{\Gamma} \hat{\sigma}^-_k$, and ensures consistent normalization and phase rotation between the measured and simulated moments. 
For simplicity, we use $\hat{a}_k\equiv \sqrt{\Gamma} \hat{\sigma}^-_k$ to denote the modes without coherent background.

\subsection{Two-mode moments calculation}
\label{Mehods-twomode}
After removing the coherent background from both propagating modes, we calculate the moments of the two propagating modes, $\hat{a}_k$, up to the fourth order. The recalculated moments
 $ \langle (\hat{a}_1^{\dagger})^{m_1} \hat{a}_1^{m_1} (\hat{a}_2^{\dagger})^{m_2} \hat{a}_2^{n_2}\rangle$ satisfy $m_1, n_1, m_2, n_2 \in \left \{  0, 1, 2\right \}$ and 
$m_1+n_1+m_2+n_2 \le 4$. These moments are computed as follows~\cite{eichlor_thesis} in Eq.~(\ref{cal_moments}),

\begin{widetext}
\begin{multline}
\langle(\hat{S_1}^{\dagger})^{m_1} \hat{S_1}^{n_1} (\hat{S_2}^{\dagger})^{m_2} \hat{S_2}^{n_2}\rangle=\sum_{i_1, j_1, i_2,j_2 =0}^{m_1, n_1, m_2, n_2}\left(\begin{array}{c}{m_1} \\ {i_1}\end{array}\right)\left(\begin{array}{c}{n_1} \\ {j_1}\end{array}\right) \left(\begin{array}{c}{m_2} \\ {i_2}\end{array}\right) \left(\begin{array}{c}{n_2} \\ {j_2}\end{array}\right) \\
\times \langle (\hat{a}_1^{\dagger})^{i_1} \hat{a}_1^{j_1} (\hat{a}_2^{\dagger})^{i_2} \hat{a}_2^{j_2}\rangle \langle(\hat{h}_1^{\dagger})^{m_1-i_1} \hat{h}_1^{n_1-j_1}(\hat{h}_2^{\dagger})^{m_2-i_2}\hat{h}_2^{n_2-j_2}\rangle\, .
\label{cal_moments}
\end{multline}
\end{widetext}

All moments shown in Results have the coherent background removed through the procedure in Methods~\ref{remove_background} before the calculation.

\subsection{Joint Quantum state tomography}
\label{Methods-QST}
\subsubsection{Optimization method}

In the joint quantum state tomography of the two propagating modes, we reconstruct the optimal density matrix from the moments of the two modes using two different methods: least-squares optimization~\cite{PhysRevResearch.2.042002, PhysRevApplied.18.044041} and compressed-sensing optimization \cite{shabani-prl-2011, rodi-prb-2014}. In this subsection, we discuss and compare the two methods.

Mathematically, the least-squares optimization method solves the following convex
optimization problem to find the optimal density matrix $\rho$: 
\begin{subequations}
\begin{alignat}{2}
&\!\min_{\rho}        &\qquad& \left\Vert (\overrightarrow{\mathcal{B}}-\mathcal{A} \overrightarrow{\rho})\oslash \overrightarrow{\epsilon} \right\Vert_{\ell_2} \label{obj_ls}\, ,\\
& \text{subject to}   &      & \rho \geq 0\, ,\label{obj:constraint1_ls}\\
&                     &      & \text{Tr}(\rho) = 1 \, ,\label{obj:constraint2_ls}
\end{alignat}
\end{subequations}
while the compressed-sensing optimization method is described by:
\begin{subequations}
\begin{alignat}{2}
&\!\min_{\rho}        &\qquad& \left\Vert \overrightarrow{\rho} \right\Vert_{\ell_1}\label{obj}\, ,\\
&\text{subject to}    &      & \left\Vert \overrightarrow{\mathcal{B}}-\mathcal{A} \overrightarrow{\rho} \right\Vert_{\ell_2} \leq \left\Vert \overrightarrow{\epsilon}\right\Vert_{\ell_2}\, ,\label{obj:constraint1}\\
&                     &      & \rho \geq 0\, ,\label{obj:constraint2}\\
&                     &      & \text{Tr}(\rho) = 1 \, ,\label{obj:constraint3}
\end{alignat}
\end{subequations}
where $\overrightarrow{\rho}$ in the objective function given in Eq.~(\ref{obj_ls}) and (\ref{obj}) represents the vectorized form of the density matrix.
In Eq.~(\ref{obj_ls}) and (\ref{obj:constraint1}),  $\overrightarrow{\mathcal{B}}$  is a column vector containing the experimentally measured or numerically simulated moments,  $ \langle (\hat{a}_1^{\dagger})^{m_1} \hat{a}_1^{n_1} (\hat{a}_2^{\dagger})^{m_2} \hat{a}_2^{n_2}\rangle$. For both optimizations, the matrix $\mathcal{A}$ is commonly referred to as the sensing matrix~\cite{PhysRevApplied.18.044041}, which only depends on the operator basis set ($\{ \vert i \rangle \langle j \vert \}$) and the measurement observable set ($\{ (\hat{a}_1^{\dagger})^{m_1} \hat{a}_1^{n_1} (\hat{a}_2^{\dagger})^{m_2} \hat{a}_2^{n_2} \}$).  $\overrightarrow{\epsilon}$ in Eq.~(\ref{obj_ls}) and (\ref{obj:constraint1}) quantifies the level of uncertainty in the measurement, which is defined as the standard deviation vector. We calculate the standard deviation by splitting the data into 20 segments, computing moments for each segment, and then calculating the standard deviation across the moments from all segments. The Hadamard division operator $\oslash$ represents the element-wise vector division in Eq.~(\ref{obj_ls}).
Furthermore, \( \|\cdot\|_{\ell_1} \) and \( \|\cdot\|_{\ell_2} \) represent the \(\ell_1\) and \(\ell_2\) norms, respectively. The \(\ell_1\) norm is calculated as the sum of the absolute values of the vector components, while the \(\ell_2\) norm, also known as the Euclidean norm, is calculated as the square root of the sum of the squared vector components.
The additional constraints given in  Eq.~(\ref{obj:constraint1_ls})-(\ref{obj:constraint2_ls}) and  Eq.~(\ref{obj:constraint2})-(\ref{obj:constraint3}) are positive semi-definite and unit trace conditions of the density matrix, ensuring its physical validity.

In the least-squares optimization, we minimize the least-squares distance between $\overrightarrow{\mathcal{B}}$ and $\mathcal{A} \overrightarrow{\rho}$ defined by \(\ell_2\) norm, weighting the different moments by dividing with
$\overrightarrow{\epsilon}$ element-wisely [Eq.~(\ref{obj_ls})]. This approach assigns lower weights in the optimization cost function to moments with higher standard deviation and larger uncertainty, and higher weights to those with less uncertainty, thereby adjusting their influence in the optimization process accordingly. Least-squares optimization is a widely used method, which can find the optimal density matrix from the measured moments [\figpanel{tomography}{a}]. 
On the other hand, compressed-sensing optimization minimizes the \(\ell_1\) norm of \(\overrightarrow{\rho}\) based on \(\rho\) being sparse---a property demonstrated in our simulation. This method is particularly advantageous when working with a heavily reduced dataset, such as the 27 moments used in our case.
By emphasizing sparsity, compressed-sensing effectively reconstructs the density matrix with fewer noisy coherent components, enabling us to see the distribution of logarithmic negativity $E_\mathcal{N}$ [\figpanel{tomography}{b}]. 
For both methods, we use convex optimization to find the minimization, using CVXPY~\cite{diamond2016cvxpy}.

\subsubsection{Details on reconstructed density matrices}

\begin{figure}
\centering
\includegraphics[width=0.95\linewidth]{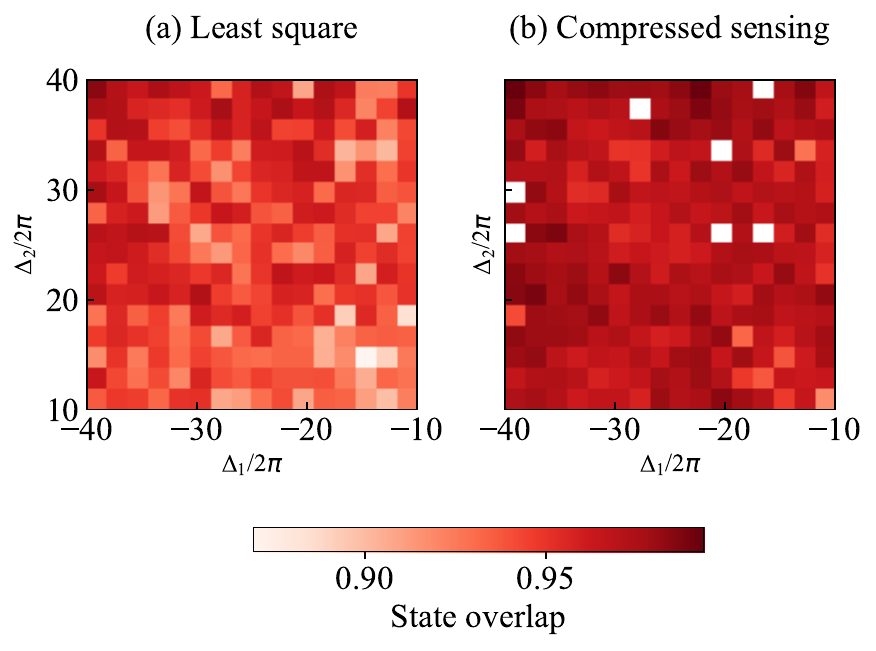}
\caption{State overlap of the joint quantum state tomography. (a) State overlap between reconstructed density matrices (using 27 measured moments) and simulated density matrices, over the frequency ranges $\Delta_1 \in [-40, -10]\,\mathrm{MHz}$ and $\Delta_2 \in [10, 40]\,\mathrm{MHz}$, utilizing least-squares optimization. (b) State overlap over the same frequency ranges, utilizing compressed-sensing optimization. Note the white points indicating not-a-number values due to optimization failures during reconstruction [see explanation under \figpanel{tomography}{b}].}
\label{state-overlap}
\end{figure}

The state overlap between the reconstructed density matrix $\rho'$ and the simulated density matrix is defined $\rho$ as~\cite{jozsa1994fidelity}, $F(\rho, \rho') = \left(\mathrm{Tr}\sqrt{\sqrt{\rho}\rho'\sqrt{\rho}}\right)^2$. Here we present the state overlap across the frequencies of the two photonic modes, for both optimizations (Fig.~\ref{state-overlap}). We observe a higher state overlap when using compressed-sensing optimization, which results in a clearer representation of logarithmic negativity compared to other methods.

\begin{figure}
\centering
\includegraphics[width=0.98\linewidth]{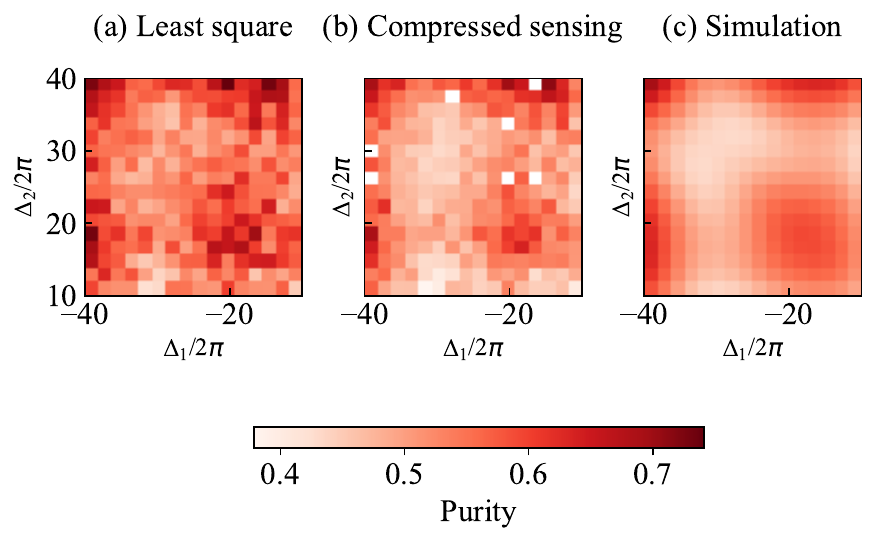}
\caption{Purity of density matrices. (a) Purity obtained from the reconstructed density matrices using 27 measured moments, utilizing least-squares optimization, over the frequency ranges $\Delta_1 \in [-40, -10]\,\mathrm{MHz}$ and $\Delta_2 \in [10, 40]\,\mathrm{MHz}$. (b) Purity obtained from the same moments as in (a), but with compressed-sensing optimization. Note the white points indicating not-a-number values due to optimization failures during reconstruction [see explanation under \figpanel{tomography}{b}]. (c) Purity obtained from numerical simulations.}
\label{purity}
\end{figure}

Furthermore, we calculate and present the purity of the density matrices at various frequencies of the two modes. The purity $P$ of a density matrix $\rho$ is defined as
\begin{equation}
    P = \text{Tr}(\rho^2)\, .
\end{equation}
We illustrate the purity of the density matrices obtained from both experimental reconstructions and numerical simulations in Fig.~\ref{purity}.

\subsubsection{Reconstruction from the simulated moments}
\label{reconst-simu}
\begin{figure}
\centering
\includegraphics[width=0.98\linewidth]{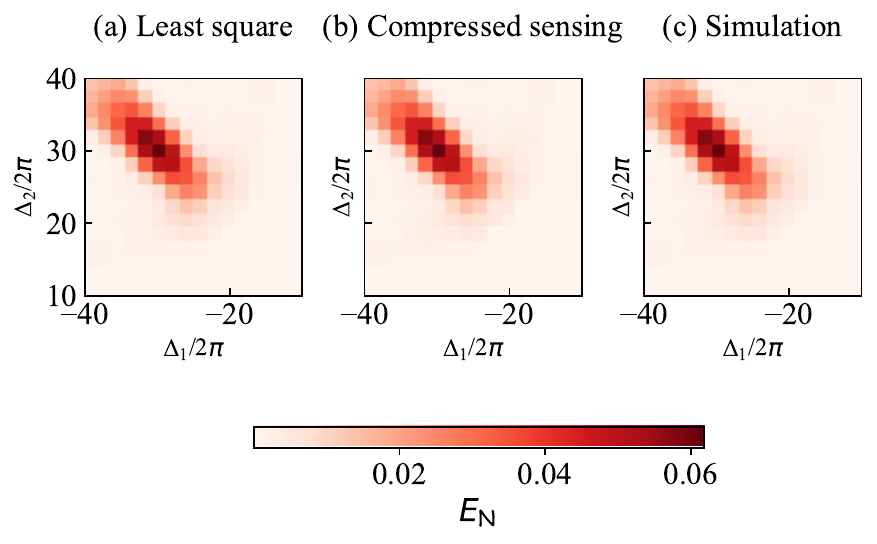}
\caption{The logarithmic negativity $E_\mathcal{N}$. (a) $E_\mathcal{N}$ reconstructed using all 325 simulated moments for a Fock space cutoff at $N=5$, utilizing least-squares optimization, over the frequency ranges $\Delta_1 \in [-40, -10]\,\mathrm{MHz}$ and $\Delta_2 \in [10, 40]\,\mathrm{MHz}$. (b) $E_\mathcal{N}$ reconstructed using the same simulated moments as in (a), but with compressed-sensing optimization, over the same frequency range. (c) $E_\mathcal{N}$ obtained from numerical simulation, over the same frequency range.}
\label{log neg simu}
\end{figure}

In Section~\ref{sec-2D}, we compare the logarithmic negativity $E_\mathcal{N}$ derived from the reconstructed density matrix using the measured moments with that obtained from simulated moments. \fullfigpanel{2D}{c} presents $E_\mathcal{N}$ calculated from 27 moments with order up to four. Here, we extend this calculation to $E_\mathcal{N}$ obtained from reconstructed density matrices using all 325 noiseless simulated moments $ \langle (\hat{a}_1^{\dagger})^{m_1} \hat{a}_1^{n_1} (\hat{a}_2^{\dagger})^{m_2} \hat{a}_2^{n_2}\rangle$ for ${m_1}$, ${m_1}$, ${m_2}$, ${n_2} \in \{0, 1, 2, 3, 4\}$ [\figpanel{log neg simu}{a} uses least-squares optimization and \figpanel{log neg simu}{b} uses compressed-sensing optimization], excluding conjugation redundancy. 
In both optimizations, we recover the same distribution as $E_\mathcal{N}$ obtained from the numerical simulation [\figpanel{log neg simu}{c}].

\section{Data Availability}
Data is available from the corresponding author upon reasonable request.

\section{Code Availability}
Code for analyzing data and generating plots is available from the corresponding author upon reasonable request.

\section{Acknowledgments}
The authors thank Janka Biznárová and Daniel Perez Lozano for help with fabrication, Kazi Rafsanjani Amin for technical assistance and useful discussions, Lars Jönsson for his help in making the sample holders, and Giulia Ferrini for her input regarding the reviewer comments. The device in this work was fabricated in Myfab, Chalmers, a micro- and nano-fabrication laboratory. The traveling-wave parametric amplifier (TWPA) used in this experiment was provided by IARPA and Lincoln Labs. This work was supported by Ericsson Research and the Knut and Alice Wallenberg Foundation through the Wallenberg Centre for Quantum Technology (WACQT). S.G. acknowledges financial support from the European Research Council via Grant No. 101041744 ESQuAT. A.F.K. is also supported by the Swedish Research Council (grant number 2019-03696), the Swedish Foundation for Strategic Research (grant numbers FFL21-0279 and FUS21-0063), and the Horizon Europe programme HORIZON-CL4-2022-QUANTUM-01-SGA via the project 101113946 OpenSuperQPlus100.
A.V.V and C.S.M. acknowledge support from the Proyecto Sin\'ergico CAM 2020 Y2020/TCS-
6545 (NanoQuCo-CM), and MCINN projects PID2021-126964OB-I00  and TED2021-130552B-C21 and PID2021-127968NB-I00.  C. S. M. also acknowledge the support of a fellowship from la Caixa Foundation (ID 100010434), from the European Union's Horizon 2020 Research and Innovation Programme under the Marie Sklodowska-Curie Grant Agreement No. 847648, with fellowship code LCF/BQ/PI20/11760026.


\section{Author contributions}
S.G. and C.S.M planned the project. J.Y. performed the experiments and analyzed the data with input from A.M.E., I.S., S.G., and M.A.U.. A.V.V., C.S.M., and I.S. built the theoretical model. J.Y. designed the device and C.C.M. helped with the fabrication of the device.  A.G. and A.F.K. operated the quantum state tomography. J.Y., I.S., and A.V.V. wrote the manuscript with feedback from all authors. S.G. and A.M.E. supervised this work. 

\section{Competing Interests}
The authors declare that they have no conflict of interest.



%

\end{document}


\setcounter{figure}{0} 
\renewcommand{\thefigure}{S\arabic{figure}} 
\renewcommand{\theequation}{S\arabic{equation}}

\newcommand{\bluetext}[1]{\textcolor{blue}{#1}}
\newcommand{\figpanel}[2]{Fig.~\hyperref[#1]{\ref*{#1}(#2)}}
\newcommand{\figpanels}[3]{Fig.~\hyperref[#1]{\ref*{#1}(#2-#3)}}
\newcommand{\figpanelNoPrefix}[2]{\hyperref[#1]{\ref*{#1}(#2)}}
\newcommand{\fullfigpanel}[2]{Figure~\hyperref[#1]{\ref*{#1}(#2)}}
\newcommand{\fullfigpanels}[3]{Fig.~\hyperref[#1]{\ref*{#1}(#2, #3)}}

\author{
Jiaying Yang$^{1,2}$,
Ingrid Strandberg$^{1}$,
Alejandro Vivas-Viaña$^{3,4}$,
Akshay Gaikwad$^{1}$,
Claudia Castillo-Moreno$^{1}$,
Anton Frisk Kockum$^{1}$,
Muhammad Asad Ullah$^{2}$,
Carlos Sánchez Muñoz$^{3,4}$,
Axel Martin Eriksson$^{1}$,
Simone Gasparinetti
}

\address{
Department of Microtechnology and Nanoscience, Chalmers University of Technology, SE-412 96, G\"{o}teborg, Sweden \\
$^2$Ericsson Research, Ericsson AB, SE-164 83, Stockholm, Sweden \\
$^3$Departamento de Física Teórica de la Materia Condensada and Condensed Matter Physics Center (IFIMAC), Universidad Autónoma de Madrid, 28049 Madrid, Spain \\
$^4$Institute of Fundamental Physics IFF-CSIC, Calle Serrano 113b, 28006 Madrid, Spain
}

\title{Entanglement of photonic modes from a continuously driven two-level system --- Supplementary Material} 
\maketitle

\section{Measurement details}
\label{setup}

\begin{figure*}[ht]
\centering
\includegraphics[width=0.75 \linewidth]{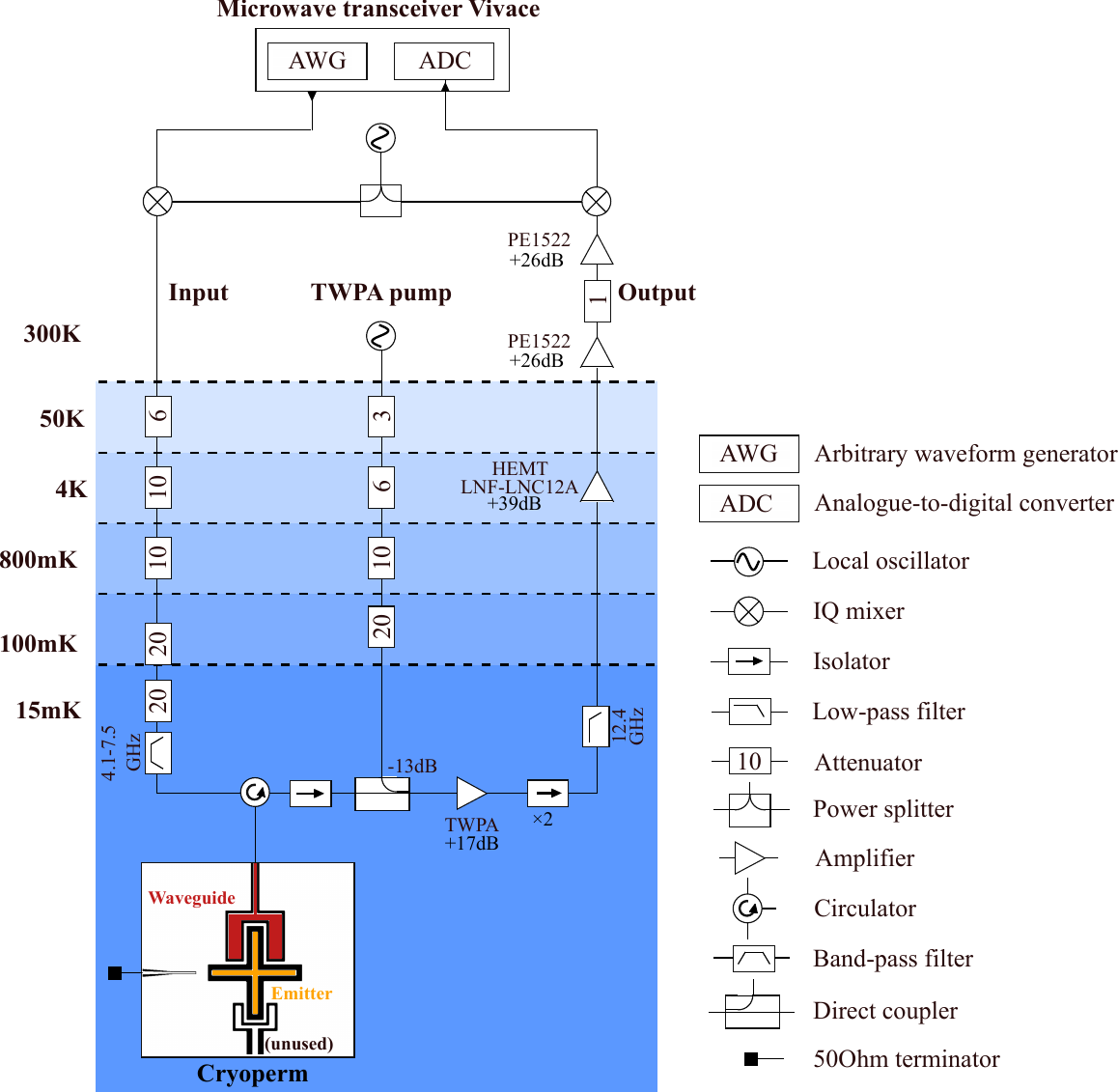}
\caption{Wiring diagram of the system, showing how the superconducting chip is connected to the input line and output line of the experimental setup. The drive signal to the emitter is applied through the input line, reaches the emitter through the coplanar waveguide, and the output signal is measured through the output line of the system. Note that the output from the AWG and the input to the ADC has two ports, including the in-phase component (I) and the out-of-phase component (Q), but only one port is shown in the figure for simplicity. The microwave transceiver board Vivace~\cite{Vivace} provides both AWG and ADC channels.}
\label{wiring diagram}
\end{figure*}

In this work, we utilize a $6.6 \times \SI{6.6}{mm^2}$ superconducting device. The device is fabricated on a silicon substrate, and its RF lines and ground plane consist of aluminium layers deposited on top of the substrate. We wire-bond the device in a copper sample holder, which is then enclosed within a copper shield. To provide additional protection against magnetic interference, we place the bonded device inside another $\mu$-metal shield (cryoperm). This shielding is installed in the mixing chamber of a dilution refrigerator to ensure that all experimental measurements are conducted at temperatures below 15 mK (Fig.~\ref{wiring diagram}).

We use the same device as in our previous work~\cite{yang2023deterministic} in this experiment. However, only the emitter qubit (orange) and the waveguide (red) are used in this work. The anharmonicity of the emitter is $\alpha/2\pi = \SI{220}{MHz}$. The unused coupler (not shown) on the device is coupled to the qubit, but has no impact due to its high frequency, \SI{7.735}{GHz}, far away from the emitter. The waveguide capacitively coupled to the emitter connects to the reflection input and output lines on the other side, from where the emitted photon field is measured via a traveling-wave parametric amplifier (TWPA)~\cite{macklin2015near} and a high-electron-mobility transistor (HEMT) amplifier in the output line. In our system, we measure the quantum efficiency $\eta$ to be 0.043, where $\eta = \frac{1/2}{1/2+n_{\rm added}}$ and $n_{\rm added}=11$ is the added noise photon number by the amplification chain. Aside from the coplanar waveguide, the emitter also capacitively couples to a charge line, which is grounded and not used in this work. 

In the experiment, data are obtained using a pulsed setup. Microwave control pulses are sent to drive the emitter using arbitrary waveform generators (AWGs), and the data are read out with analogue-to-digital converters (ADCs) from the microwave transceiver platform Vivace~\cite{Vivace}, after up- and down-conversion by IQ mixers and local oscillators. Temporal mode matching and data acquisition begin when the qubit drive has lasted for $t_0$ = \SI{200}{ns}---a duration much longer than the qubit's relaxation time $T_1$---ensuring the qubit is in a steady state.

\section{Theoretical results on entanglement generation}
\label{simulation results}
\begin{figure*}
\centering
\includegraphics[width=\linewidth]{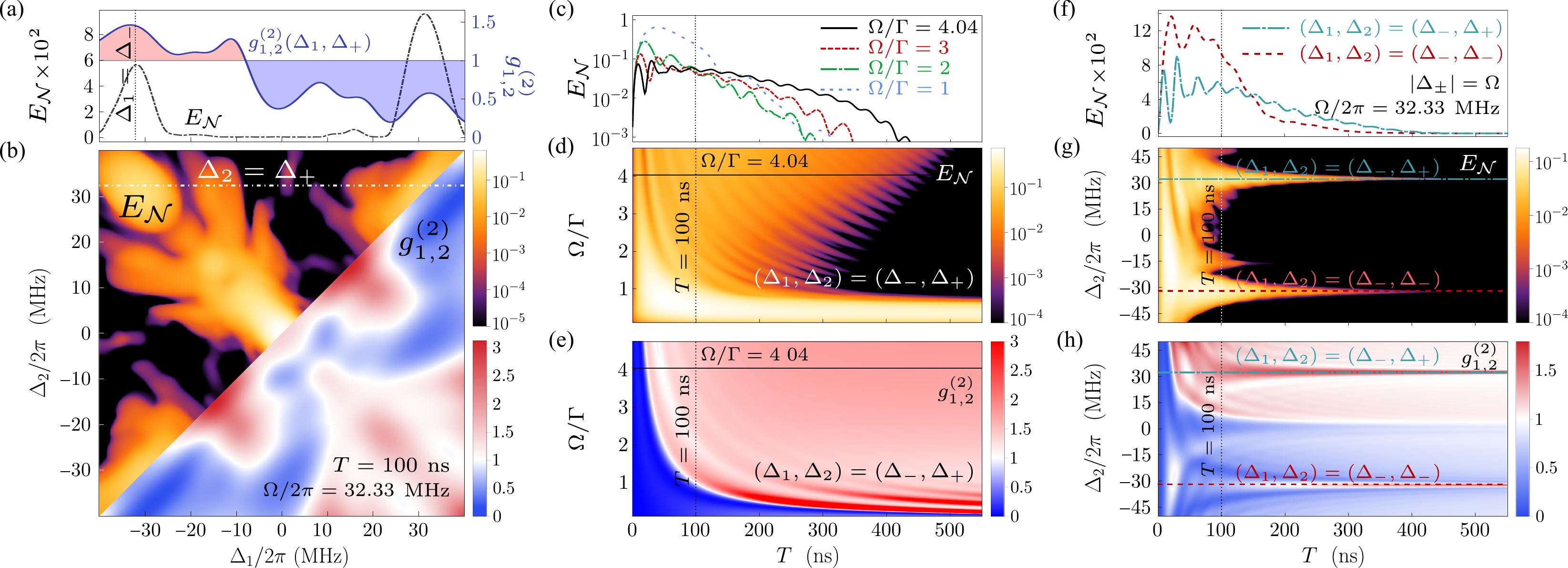}
\caption{
Simulated logarithmic negativity (in logarithmic scale), $E_\mathcal{N}$, and two-photon correlation function at zero delay, $g_{1,2}^{(2)}$. 
(a) $E_\mathcal{N}$ (black dotted-dashed line) and $g_{1,2}^{(2)}(\Delta_1,\Delta_+)$ (blue solid line) in terms of the filter detuning $\Delta_1$ [$\Delta_2=\Delta_+$, white dotted-dashed line in (b)]. The vertical dotted line denotes the resonant frequency, $\Delta_1=\Delta_-$.
(b) $E_\mathcal{N}$ (upper-left triangle) and $g_{1,2}^{(2)}$ (lower-right triangle) in the frequency domain. Both figures are symmetric with respect the diagonal.
(c) $E_\mathcal{N}$ at opposite sidebands [$(\Delta_1,\Delta_2)=(\Delta_-,\Delta_+)$, equivalently in (d,e)] in terms of the measurement time $T$ for several pulse intensities $\Omega=\{1,2,3,4.04\}\Gamma$. The latter is actually used in the experiment. 
(d,e) $E_\mathcal{N}$ and $g_{1,2}^{(2)}$, respectively, in terms of Rabi frequency of the drive $\Omega$ and the measurement time $T$. In (e), for practical reasons, the data is upper-bounded up to $3$. The actual maximum value, $\approx 36$, only occurs at low pumping and long filtering times (bottom-right red areas). 
(f) $E_\mathcal{N}$ in terms the measurement time $T$ for the two sideband resonances, $(\Delta_1,\Delta_2)=(\Delta_-,\Delta_\pm)$, in blue dot-dashed and red dashed lines, respectively [equivalently in (g,h)].
(g,h) $E_\mathcal{N}$ and $g_{1,2}^{(2)}(\Delta_-,\Delta)$, respectively, in terms of the filter detuning $\Delta_2$ and the measurement time $T$.
The vertical dot-line and horizontal solid line in (c--h) correspond to the experimental values used in the experiment: $(\Omega,T)=(32.32\ \text{MHz},100 \ \text{ns})$.
Parameters: $\Gamma/2\pi=8$ MHz, $\Omega/2\pi=32.33$ MHz, $\Delta=0$, $t_0=200$ ns, $T=100$ ns. }
\label{fig:negativity}
\end{figure*}
In this section, we extend Results by simulating the logarithmic negativity and the two-photon correlation function at zero delay in the frequency domain. We study the effects of varying the driving strength of the microwave pulse ($\Omega$), the frequency of the filters ($\Delta_k$), and the duration of the measurement ($T$) [see \figpanels{fig:negativity}{a}{h}]. 

\subsection{Entanglement in the frequency domain}

Throughout this work we focused on the specific case where the output modes correspond to opposite sidebands. However, by extending the range of frequencies in the filtering process, we unveil a rich structure in the generation of entanglement, as depicted in \fullfigpanels{fig:negativity}{a}{b}. 
In the frequency map [upper-left triangle in \figpanel{fig:negativity}{b}], we observe two main regions of non-zero logarithmic negativity: the diagonal ($\Delta_1=\Delta_2$) and the antidiagonal ($\Delta_2=-\Delta_1$). However, only the later is relevant to this work, as discussed below. 

The diagonal section entails a breakdown of the orthogonality assumption between modes, which we consider a basic requirement for discussing entanglement between two different systems.
More generally, this assumption will be compromised in sections where $|\Delta_1-\Delta_2|<\Delta \omega$ or $\Omega < \Gamma$---where $\Delta \omega$ is the frequency resolution inherited by the filter---since the reported values of entanglement cannot be understood as actual measures of entanglement. In these cases, spectral resolution is lost and the collected modes become indistinguishable in frequency.  
Nevertheless, this information reveals the amount of entanglement achievable if the whole $a_\text{out}$ is physically split into two orthogonal modes (e.g., by a beam splitter) prior to filtering, in which case entanglement between any two pair of temporal modes extracted from each of the split outputs would be well-defined. As discussed in Methods A, this requires a slightly different master equation accounting for the vacuum contributions introduced by splitting the signal, with results strongly correlated to those presented here (not shown).

Therefore, regimes in which $\Omega >\Gamma$ and $|\Delta_1-\Delta_2|>\Delta \omega$ are the most interesting since it provides a well-resolved emission in different spectral lines, resulting in well-defined entanglement. 

Additionally, in the lower panel in \figpanel{fig:negativity}{b}, we observe the photon-photon correlations of the filtered emission 
by computing the zero-delay two-photon cross-correlation
\begin{equation}
    g_{1,2}^{(2)}=\frac{\langle  \hat a_1^\dagger  \hat a_2^\dagger \hat a_1 \hat a_2\rangle}{\langle \hat a_1^\dagger \hat a_1\rangle \langle \hat a_2^\dagger \hat a_2\rangle}.
\end{equation}
As discussed in Ref.~\cite{Gonzalez-TudelaTwophotonSpectra2013}, this map exhibits the typical frequency structure of a coherently driven TLS, where the antidiagonal features a complex structure of correlations due to the multi-photon processes occurring at these frequencies. In fact, in \figpanel{fig:negativity}{a}, we observe a clear correspondence between entanglement and bunching around the frequency resonance, specifically at $(\Delta_1,\Delta_2)=(\Delta_-,\Delta_+)$. 

\begin{figure*}[h]
\centering
\includegraphics[width=0.55\linewidth]{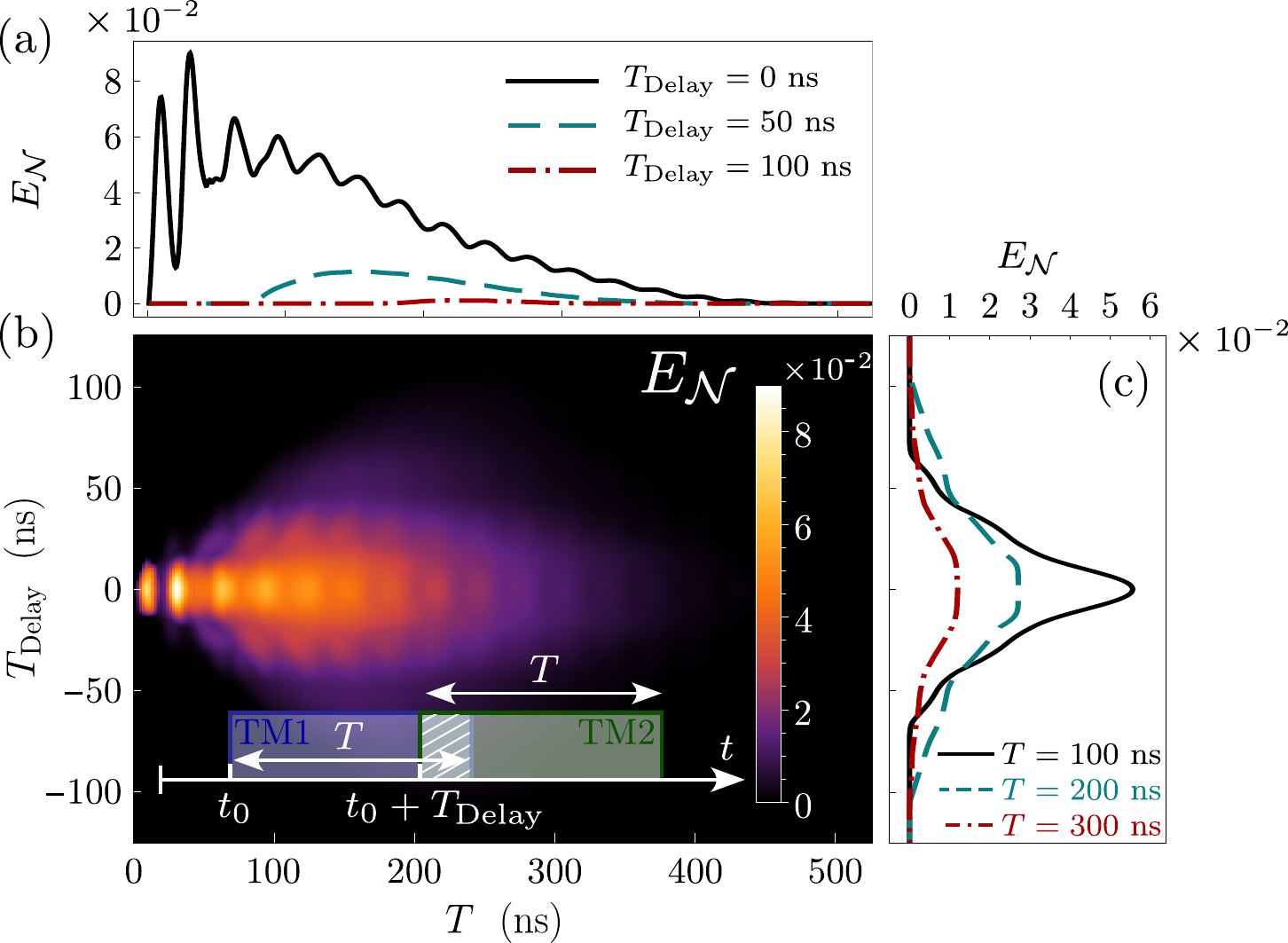}
\caption{
Simulated logarithmic negativity $E_\mathcal{N}$, evaluated as a function of the time delay between two temporal boxcar filters and their durations. (a) Map of $E_\mathcal{N}$ as a function of both starting-time delay $T_\textrm{Delay}$ and duration $T$. The inset illustrates the time sequence of the applied temporal filters: the first filter, TM1, starts at a fixed time $t_0$, while the starting time of the second filter, TM2, is varied. (b) Dependence of $E_\mathcal{N}$ on $T$ for selected values of $T_\textrm{Delay}$. (c) Dependence of $E_\mathcal{N}$ on $T_\textrm{Delay}$ for selected values of $T$. Entanglement decreases as the delay is increased.}
\label{fig:T_delay}
\end{figure*}

Throughout this discussion, we have assumed that the two-mode temporal profiles, $v_1(t)$ and $v_2(t)$, are perfectly overlapping in time, while their frequencies remain independently tunable. Here, we investigate how relaxing the requirement of perfect temporal overlap influences the entanglement. Specifically, we analyze a scenario in which the boxcar-shaped temporal filters applied to the modes are temporally offset from each other, introducing a controllable time delay $T_\textrm{Delay}$ between them. We perform simulations where the starting time of the filter for the second temporal mode (TM2) is adjusted to \( t_0 + T_\textrm{Delay} \), with \( T_\textrm{Delay} \) varied within the range \([-0.125, 0.125] \, \mu\mathrm{s}\) (Fig.~\ref{fig:T_delay}). The initial starting time of the other mode (TM1) is fixed at \( t_0 \). Additionally, the duration of the templates is systematically varied to examine its influence. The logarithmic negativity $E_{\mathcal{N}}$, measured at the maximum-entanglement point $(\Delta_{1}, \Delta_{2}) = (\Delta_{-}, \Delta_{+})$, exhibits a symmetric decaying pattern centered around $T_{\mathrm{Delay}} = 0$, implying that for any duration, entanglement is maximized by choosing the two modes to perfectly overlap in time. 

\subsection{Effects of the filtering time}

The dependence on $T$ in all the results in \figpanels{fig:negativity}{c}{h} exhibits similar behaviour. The boxcar filter imposes an approximate bandwidth of $\Delta\omega\sim 2\pi/T$---corresponding to the width of the central peak of the sinc function---such that in the limit of infinite frequency precision, $T\rightarrow \infty$, the filtered modes lose their time definition, causing entanglement to vanish.
Conversely, very short measurement times, $T\rightarrow 0$, correspond to broadband detection (colour-blind) [see Fig.~2(b) in Results], resulting in a non-zero logarithmic negativity that does not reflect an actual measure of entanglement, as we previously discussed. Additionally, in this limit, we recover the expected antibunching of a two-level system.
We also note that higher values of $\Omega$ extend the lifetime of the entanglement [see \fullfigpanels{fig:negativity}{c}{d})]. 

In the limit of long $T$ (i.e., very narrow frequency filtering), we observe that the photon statistics is not completely uncorrelated, $\lim_{T\rightarrow \infty}g_{1,2}^{(2)}(\Delta_-,\Delta_\pm)\neq 1$, but bunched ($\lesssim 2$) [see  \fullfigpanels{fig:negativity}{e}{h}].
This feature was already reported in Refs.~\cite{AspectTimeCorrelations1980,DalibardCorrelationSignals1983,BelTheoryWavelengthResolved2009,Gonzalez-TudelaTwophotonSpectra2013}, where it is shown that correlations between opposite sidebands feature bunching when a perfect laser is used to drive the TLS. To recover the expected limit of uncorrelated photons when $\Delta_1\neq \Delta_2$, a more realistic model of a laser must be considered, such as an one-atom laser~\cite{MuOneatomLasers1992}.
In \fullfigpanels{fig:negativity}{f}{g}, we see that when one of the filters is fixed on a Mollow sideband, e.g., $\Delta_1=\Delta_-$, the collected emission by the other filter features two main resonances that contribute to the generation of entanglement. These resonances correspond to (i) opposite sidebands, $(\Delta_1,\Delta_2)=(\Delta_-,\Delta_+)$ (in blue), and (ii) identical sidebands, $(\Delta_1,\Delta_2)=(\Delta_-,\Delta_-)$ (in red). However, as previously discussed, the nature of these entangled states differs since only the first case involves entanglement between two distinct, well-resolved light modes.
%

\subsection{Entanglement in the time domain}

\begin{figure*}[h]
    \centering
\includegraphics[width=0.4\linewidth]{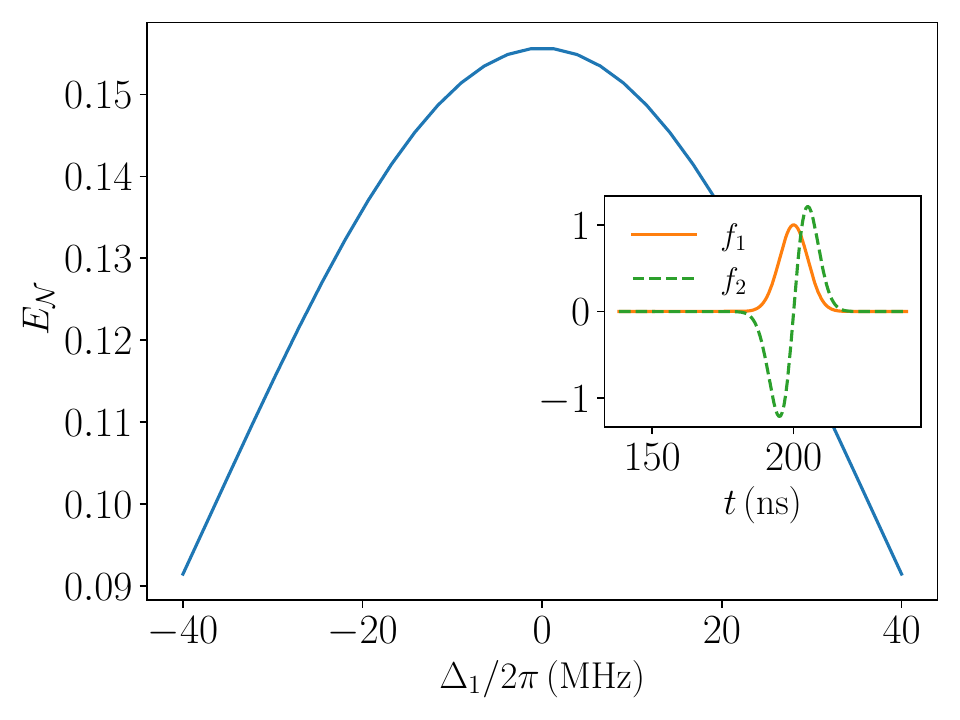}
\caption{Simulated logarithmic negativity for the first two Hermite--Gauss modes $n=0,1$ in Eq.~\eqref{eq:hermite_gauss} with $w=0.5$. The detuning $\Delta_1$ of the first mode is varied while $\Delta_2=0$ is fixed. The entanglement is strongest when there is no detuning. Inset: The Hermite--Gauss modes in the time domain centered at $t_0$. }
\label{fig:logNeg_temporalModes}
\end{figure*}

Besides orthogonal frequency modes with a boxcar temporal filter, there are other options for the mode shapes. A common temporal mode basis is the \emph{Hermite--Gauss} modes. They have certain overlap in both time and frequency, but are field-orthogonal~\cite{Brecht2015Oct, RaymerTemporalModes2020}. The mode shape is defined as
\begin{equation}\label{eq:hermite_gauss}
    f_i(t) = H_n\left(\frac{t}{w}\right) \exp\left(-\frac{t^2}{2w^2}\right),
\end{equation}
where $H_n$ is the $n$th Hermite polynomial, and $w$ determines the temporal width of the mode. Index $i=1,2$ indicates our previous mode filter numbering. We simulate logarithmic negativity for Hermite--Gauss modes with $n=0$ and $n=1$ using $w=0.5$ in Fig.~\ref{fig:logNeg_temporalModes}. The maximum logarithmic negativity ($E_\mathcal{N}=0.155$) is more than twice as large as that of the boxcar modes ($E_\mathcal{N}=0.062$). This improvement in logarithmic negativity indicates potential advantages of exploring alternative mode shapes beyond boxcar modes. 

\section{TWPA gain compensation}
\begin{figure*}[ht]
\centering
\includegraphics[width=0.95 \linewidth]{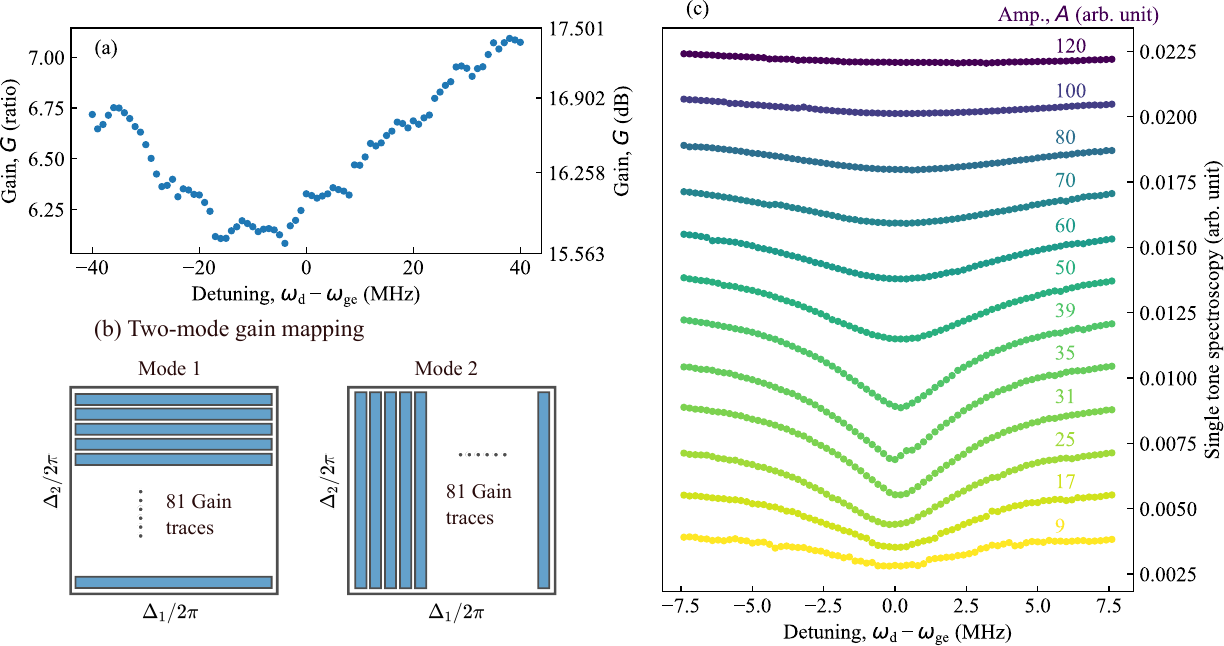}
\caption{TWPA gain and AWG amplitude calibration for the system. (a) The on-off gain of the TWPA as a function of the detuning between the drive frequency $\omega_{\rm d}$ and the qubit frequency $\omega_{\rm ge}$, presented in ratio and with unit dB. (b) Schematic diagram showing how we map the gain trace in (a) into gain matrices, for two-mode data compensation. The same time trace is repeated horizontally (vertically) for mode 1 (2) data $\hat{a}_1$ ($\hat{a}_2$). Each gain matrix contains 81 gain traces, corresponding to the number of points along the frequency detuning axes. (c) The single-tone spectroscopy of the qubit measured with ADC, when applying microwave pulses with different driving amplitude from AWG. Each trace is normalized with the corresponding driving amplitude and is offset by 0.0017 compared to the trace below it. }
\label{calibration}
\end{figure*}

To calibrate the gain of the TWPA, we conduct a single-tone spectroscopy analysis of the qubit through the waveguide. This is achieved under the condition that the qubit is saturated by utilizing high power in the microwave input pulse. The spectroscopy is performed twice across the bandwidth of $[-40, 40]$ MHz around the qubit's resonance frequency, both with the TWPA powered and unpowered. We obtain the on-off gain of the TWPA by calculating the ratio between the spectroscopy measured in the two cases [\figpanel{calibration}{a}]. 

Due to the observed variation in TWPA gain across the specified bandwidth, we apply compensation to both temporal-matched one-mode and two-mode data of the propagating modes. One-mode data (see Fig.~2 in Results) is compensated by dividing the data with the gain trace in  \figpanel{calibration}{a}. For two-mode data (see Fig.~3 in Results) which are two-dimensional matrices for both modes $\hat{a}_1$ and $\hat{a}_2$, we map the gain trace into gain matrices according to \figpanel{calibration}{b}, allowing us to compensate $\hat{a}_1$ and $\hat{a}_2$ by dividing the corresponding gain matrices. 
\section{Conversion between AWG amplitude to Rabi rate}
We obtain the conversion between the amplitude of the AWG output pulse and the drive Rabi frequency $\Omega$ by operating power calibration of the qubit [\figpanel{calibration}{c}]. We drive the qubit with pulses from the AWG with different amplitudes, and measure the single-tone spectroscopy of the qubit from the reflection setup connected to the waveguide. At the critical power of the qubit, where the qubit-driving power maximizes the coherence between the ground and excited states, the strength of the input pulse is $\Omega = \Gamma/\sqrt{2}$ \cite{scigliuzzo2020primary}. From the measurement, we observe that when the driving amplitude is $A_{\rm critic} = 35$ (arb. unit), the qubit spectroscopy has the largest dip, corresponding to the case with the critical power. Due to the proportional relationship between the drive rate $\Omega$ and the  driving amplitude $A_{\rm in}$, any strength can therefore be converted from the input power $P_{\rm in}$ of the driving pulse by following 
\begin{equation}
\Omega = \frac{A_{\rm in}}{A_{\rm critic}}  \frac{\Gamma}{\sqrt{2}} \, .
\end{equation}

\nocite{*}

%